\documentclass[conference]{IEEEtran}

\usepackage{amsmath,amssymb,amsfonts}
\usepackage{bm}
\usepackage{booktabs}
\usepackage{cite}
\usepackage[T1]{fontenc}
\usepackage{graphicx}
\usepackage{multirow}
\usepackage{siunitx}
\usepackage{soul}
\usepackage{textcomp}
\usepackage[color=green!25]{todonotes}
\soulregister\cite7
\soulregister\footnote7
\soulregister\eqref7
\soulregister\ref7
\usepackage{xcolor}
\usepackage{url}
\usepackage{hyperref}
\graphicspath{{./}{figure/}{../figure/}}

\begin{document}

\title{%
  Multi-Output Gaussian Process-Based Data Augmentation for Multi-Building and
  Multi-Floor Indoor Localization
}

\author{%
  \IEEEauthorblockN{Zhe Tang\IEEEauthorrefmark{1}\IEEEauthorrefmark{2}, Sihao
    Li\IEEEauthorrefmark{1}\IEEEauthorrefmark{2}, Kyeong Soo
    Kim\IEEEauthorrefmark{1}, and Jeremy Smith\IEEEauthorrefmark{2}}%
  \IEEEauthorblockA{\IEEEauthorrefmark{1}School of Advanced Technology,
    Xi'an Jiaotong-Liverpool University, Suzhou, China.\\
    Email:
    Zhe.Tang15@student.xjtlu.edu.cn,~Sihao.Li19@student.xjtlu.edu.cn,~Kyeongsoo.Kim@xjtlu.edu.cn}%
  \IEEEauthorblockA{\IEEEauthorrefmark{2}Electrical Engineering and Electronics,
    University of Liverpool, Liverpool, U.K.\\
    Email: \{Zhe.Tang,~Sihao.Li,~J.S.Smith\}@liverpool.ac.uk}%
}%

\maketitle

\begin{abstract}
  Location fingerprinting based on Received Signal Strength Indicator (RSSI)
  becomes a mainstream indoor localization technique due to its advantage of not
  requiring the installation of new infrastructure and the modification of
  existing devices, especially given the prevalence of Wi-Fi-enabled devices and
  the ubiquitous Wi-Fi access in modern buildings. The use of Artificial
  Intelligence (AI)/Machine Learning (ML) technologies like Deep Neural Networks
  (DNNs) makes location fingerprinting more accurate and reliable, especially
  for large-scale multi-building and multi-floor indoor localization. The
  application of DNNs for indoor localization, however, depends on a large
  amount of preprocessed and deliberately-labeled data for their
  training. Considering the difficulty of the data collection in an indoor
  environment, especially under the current epidemic situation of COVID-19, we
  investigate three different methods of RSSI data augmentation based on
  Multi-Output Gaussian Process (MOGP), i.e., \textit{by a single floor},
  \textit{by neighboring floors}, and \textit{by a single building}; unlike
  Single-Output Gaussian Process (SOGP), MOGP can take into account the
  correlation among RSSI observations from multiple Access Points (APs) deployed
  closely to each other (e.g., APs on the same floor of a building) by
  collectively handling them. The feasibility of the MOGP-based RSSI data
  augmentation is demonstrated through experiments based on the state-of-the-art
  Recursive Neural Network (RNN) indoor localization model and the UJIIndoorLoc,
  i.e., the most popular publicly-available multi-building and multi-floor
  indoor localization database, where the RNN model trained with the
  UJIIndoorLoc database augmented by using the whole RSSI data of a building in
  fitting an MOGP model (i.e., by a single building) outperforms the other two
  augmentation methods as well as the RNN model trained with the original
  UJIIndoorLoc database, resulting in the mean three-dimensional positioning
  error of \SI{8.42}{\m}.
\end{abstract}

\begin{IEEEkeywords}
  Indoor localization, data augmentation, multi-output Gaussian process (MOGP),
  recurrent neural network (RNN).
\end{IEEEkeywords}

\section{Introduction}
\label{sec:introduction}
Triangulation-based indoor localization methods like Time of Arrival (ToA), Time
Difference of Arrival (TDoA), and Angle of Arrival (AoA) have been extensively
studied for decades and well known for their sub-meter-level localization
accuracy~\cite{survey}. However, the stricter requirements of the clock
synchronization among all transmitters and/or receivers in a system for ToA/TDoA
and the use of antenna arrays for AoA not suitable for small devices, in
addition to their dependence on anchor nodes whose exact locations are to be
known, make it impractical to deploy them in large-scale multi-building and
multi-floor indoor environments. Location fingerprinting based on Received
Signal Strength Indicator (RSSI) or geomagnetic field intensity, on the other
hand, becomes a mainstream indoor localization technique due to its advantage of
not requiring the installation of new infrastructure and the modification of
existing devices~\cite{zhenghang18:_xjtluin}.

As Artificial Intelligence (AI)/Machine Learning (ML) technologies become mature
and wide-spread, the use of AI/ML for location fingerprinting is now one of the
hottest research topics, which is demonstrated, for instance, through the use of
deep neural networks (DNNs) for large-scale multi-building and multi-floor
indoor localization based on not only feedforward neural networks
(FNNs)~\cite{DNN} but also advanced DNNs like convolutional neural networks
(CNNs)~\cite{CNN1,CNN2} and recursive neural networks
(RNNs)~\cite{2021hierarchical}.
Note that the application of DNNs for indoor localization depends on a large
amount of preprocessed and deliberately-labeled data for their training. Because
the data collection in an indoor environment, especially under the current
epidemic situation of COVID-19, is costly in terms of time and man-power, the
importance of publicly-available datasets cannot be stressed enough for the
research and development of indoor localization schemes.

The \textit{UJIIndoorLoc} database from the Jaume~I~University in Spain is one
of the most popular publicly-available datasets and provides multi-building and
multi-floor Wi-Fi RSSI data, which were collected over three buildings with four
to five floors~\cite{UJI}.
Although the UJIIndoorLoc is among the largest Wi-Fi RSSI dataset with various
features, the coverage of measurement places---called \textit{reference
  points}---is limited to part of the areas: For example, the number of
reference points for the third floor of building~2 is 2709, while this number is
significantly reduced to 948 for the third floor of building~1, which might have
been related with the measurement strategy. Our investigation shows that the
uneven coverage of reference points in the UJIIndoorLoc dataset affects the
performance of the location estimation by DNNs due to the lack of training for
less-covered areas.

Data augmentation based on the interpolation/extrapolation of existing RSSI data
could be used to mitigate the impact of the lack of coverage in location
estimation. One possible approach is \textit{Gaussian Process (GP) regression},
also called \textit{Kriging} in
geostatistics~\cite{rasmussenGaussianProcessesMachine2006}; using the existing
data, GP regression can give an optimal linear unbiased prediction at unsampled
locations. In the existing work for RSSI data augmentation (e.g.,
\cite{Kriging}), GP regression is used to exploit the spatial correlation among
RSSI observations, which, however, is limited to those from one Access Point
(AP) only, i.e., the regression based on Single-Output GP (SOGP). Note that
there could be a significant correlation among RSSI observations from multiple
APs which are deployed closely to each other (e.g., APs on the same floor of a
building). To exploit the correlation among observations from multiple APs,
therefore, we extend GP regression to the case of Multi-Output GP (MOGP) and
investigate an optimal way of RSSI data augmentation based on MOGP in this
paper.

Another data augmentation algorithm for indoor localization has been recently
proposed in~\cite{njimaIndoorLocalizationUsing2021}, which is based on a
selective Generative Adversarial Network (GAN) for the generation of fake
unlabeled RSSI data and a semi-supervised DNN for pseudo-labeling of the
generated fake RSSI data. Unlike our MOGP-based approach exploiting the
correlation among observations from APs over multiple floors, the use of
selective GAN in~\cite{njimaIndoorLocalizationUsing2021} is limited to the RSSI
data of a single floor of a building. Also, their algorithm requires two
separate neural networks for the generation of fake RSSI vectors and their
pseudo-labels, respectively, and depends on rather complicated selection
criteria to address the issue of the limited coverage of the generated fake RSSI
data because the selective GAN cannot directly model the mapping between
location coordinates and RSSI data. Our approach based on MOGP regression, on
the other hand, fits a model based on the existing RSSI data, which can be
sampled at any point in a given area; hence, it could solve the uneven spatial
distribution and lack of RSSI data relatively easily and flexibly.

\section{Multi-Output Gaussian Process (MOGP)}
\label{sec:mogp}

\subsection{Regression}
\label{sec:math}
Let $X$ and $Y$ be a training set of $L$-dimensional input points (e.g.,
location coordinates) and a set of the corresponding $M$-dimensional output
observations (e.g., RSSIs for multiple APs), which are defined as follows:
\begin{equation}
  \label{eq:inputs-observations}
  \begin{aligned}
    X & = \left\{\bm{x}_{1}, \ldots, \bm{x}_{N}\right\}, \\
    Y & = \left\{\bm{y}(\bm{x}_{1}), \ldots, \bm{y}(\bm{x}_{N})\right\},
  \end{aligned}
\end{equation}
where
\begin{equation}
  \begin{aligned}
    \bm{x}_{i} & = \left(x_{i,1},{\ldots},x_{i,L}\right)^{\top}, \\
    \bm{y}(\bm{x}_{i}) & =
    \left(y_{1}(\bm{x}_{i}),{\ldots},y_{M}(\bm{x}_{i})\right)^{\top}.
  \end{aligned}
\end{equation}
for $i{=}1,{\ldots},N$. Given $X$ and $Y$, MOGP collectively approximates $M$
outputs $\bm{f}{=}\{f_{1},{\ldots},f_{M}\}^{\top}$ based on the assumption that
their joint distribution is given by a multivariate Gaussian distribution, i.e.,
\begin{equation}
  \bm{f}(\bm{x}) \sim \mathcal{GP}\left(\bm{m}(\bm{x}),\bm{K}(\bm{x},\bm{x'})\right),
\end{equation}
where $\bm{m}(\bm{x})$ is a mean vector, i.e.,
$(m(x_{1}),{\ldots},m(x_{L}))^{\top}$, which are typically set to $\bm{0}$, and
$\bm{K}(\bm{x},\bm{x'})$ is a covariance matrix defined as follows:
\begin{multline}
  \label{eq:covariance-matrix}
  \bm{K}(\bm{x},\bm{x'}) = \\
  \begin{bmatrix}
    k_{1,1}(\bm{x},\bm{x'}) & k_{1,2}(\bm{x},\bm{x'}) & \cdots & k_{1,M}(\bm{x},\bm{x}') \\
    k_{2,1}(\bm{x},\bm{x'}) & k_{2,2}(\bm{x},\bm{x'}) & \cdots & k_{2,M}(\bm{x},\bm{x'}) \\
    \vdots         & \vdots         & \ddots & \vdots         \\
    k_{M,1}(\bm{x},\bm{x'}) & k_{M,2}(\bm{x},\bm{x'}) & \cdots & k_{M,M}(\bm{x},\bm{x'}) \\
  \end{bmatrix},
\end{multline}
where $k_{i,j}(\cdot,\cdot)$ is a covariance function also called
\textit{kernel}~\cite{intro_kernel}; unlike the mean, the kernel heavily impacts
the fitting performance of MOGP~\cite{mean}.

Note that the observations with independent and identically distributed Gaussian
measurement noise are modeled as follows:
\begin{equation}
  \label{eq:observation-model}
  y_{i} = f_{i}(x) + \epsilon_{i}, ~~ i = 1, \ldots, M,
\end{equation}
where $\epsilon_{i}{\sim}\mathcal{N}(0,\sigma^{2}_{i})$. In this case, the
likelihood function $\mathcal{L}$ is given by
\begin{equation}
  \label{eq:likelihood}
  \mathcal{L}\left(\bm{f},\bm{x},\Sigma|\bm{y}\right) =
  p(\bm{y}|\bm{f},\bm{x},\Sigma) = \mathcal{N}\left(\bm{f}(\bm{x}),\Sigma\right),
\end{equation}
where $\Sigma$ is a diagonal matrix with $\{\sigma^{2}_{i}\}_{1{\leq}i{\leq}M}$.
Also, the posterior distribution of $\bm{f_{*}}{\triangleq}\bm{f}(\bm{x_{*}})$
at a test point $\bm{x_{*}}$ given $X$ and $Y$ is
\begin{equation}
  \label{eq:post-dist}
  \bm{f_{*}}|\bm{x_{*}},X,Y \sim \mathcal{N}\left(\bm{\hat{f}}(\bm{x_{*}}),\hat{\Sigma}(\bm{x_{*}})\right),
\end{equation}
where $\bm{\hat{f}}(\bm{x_{*}})$ and $\hat{\Sigma}(\bm{x_{*}})$ are the mean and
the variance of the prediction, respectively.\footnote{The readers are referred
  to~\cite{MOGPremarks} for the details of their derivations.}
Fitting an MOGP with a covariance matrix having a general structure, by the way,
is quite difficult; the computational complexity for $n$ inputs and $p$ outputs
is $\mathcal{O}(n^{3}p^{3})$~\cite{bruinsmaScalableExactInference}. Given the
symmetric nature of RSSI data for indoor localization, therefore, we consider
symmetric MOGPs in this paper per the suggestion from~\cite{MOGPremarks}, of
which we focus on separable models treating the inputs and outputs separately.

Using the generative approach called Linear Model of Coregionalization (LMC), we
can model MOGP outputs as a linear combination of $R$ \textit{latent functions}
as follows:
\begin{equation}
  \label{eq:lmc-mogp}
  f_{i}(\bm{x}) = \sum_{r=1}^{R}\alpha_{i,r}g_{r}(\bm{x}), ~~ i = 1, \ldots, M,
\end{equation}
where $g_{r}(\bm{x}){\sim}\mathcal{GP}(0,k_{r}(\bm{x},\bm{x'}))$ and
$\alpha_{i,r}$ is its coefficient.
A special case of $R{=}1$ is called Intrinsic Coregionalization Model (ICM), and
its covariance matrix is given by
\begin{equation}
  \label{eq:icm-cm}
  \bm{K}(\bm{x},\bm{x'}) = \bm{A}k_{1}\left(\bm{x},\bm{x'}\right),
\end{equation}
where
\begin{equation}
  \bm{A} = 
  \begin{bmatrix}
    \alpha_{1,1}\alpha_{1,1} & \cdots & \alpha_{1,1}\alpha_{M,1} \\
    \vdots                  & \ddots & \vdots         \\
    \alpha_{M,1}\alpha_{1,1} & \cdots & \alpha_{M,1}\alpha_{M,1} \\
  \end{bmatrix}.
\end{equation}
%
The covariance matrix can be greatly simplified by LMC and ICM, but we still
need to specify the kernel---i.e., $k_{1}$ in~\eqref{eq:icm-cm}---for a given
application. A new kernel can be obtained by linear operations of the known
kernel functions~\cite{kernel_calculation}, and one of the most wildly used
kernels is the Square-Exponential kernel, also called Gaussian Radial Basis
Function (RBF)~\cite{RBF}:
\begin{equation}
  \label{eq:gaussian-rbf}
  k_{\gamma }(\bm{x},\bm{x'}) = \operatorname{exp}\left(-\frac{\left\| \bm{x}-\bm{x'}\right\|^{2}}{\gamma^{2}}\right).
\end{equation}
Since it is infinitely differentiable, the function becomes smoother as $\gamma$
increase; some argues in this regard that the Gaussian RBF is too smooth to
model applications in the physical world~\cite{RBF}. In this paper, therefore, we
use the Mat\'{e}rn kernel with coefficients $\alpha{=}5/2$, which is not overly
smooth~\cite{Matern}:
\begin{align}
  \label{eq:matern-kernel}
  k_{5/2,h}(\bm{x},\bm{x'}) = & \left(1+\frac{\sqrt{5}\left\|
                                \bm{x}-\bm{x'}\right\|}{h}+\frac{\sqrt{5}\left\|
                                \bm{x}-\bm{x'}\right\|^{2}}{3h^{2}}\right)
                                \nonumber \\
                              & \times \operatorname{exp}\left(-\frac{\sqrt{5}\left\|
                                \bm{x}-\bm{x'}\right\|}{h}\right).
\end{align}

\subsection{Advantages over SOGP in RSSI Data Augmentation}
\label{sec:moso}
The reason for using MOGP for RSSI data augmentation instead of SOGP is
two-fold.

First, MOGP is known to outperform SOGP in predicting multivariate outputs
because MOGP can exploit the correlation among the related outputs; for
instance, it was demonstrated through the experimental assessment of the
prediction performance of GP with eight examples that eight of the ten MOGP
models with different characteristics show superiority over
SOGP~\cite{MOGPremarks}.

Second, the very problem of RSSI data augmentation for multi-building and
multi-floor indoor localization suits well MOGP regression as shown in
Fig.~\ref{fig:gp-data-augmentation}:
\begin{figure}[!tbh]
  \begin{center}
    \includegraphics[angle=-90,width=.5\linewidth]{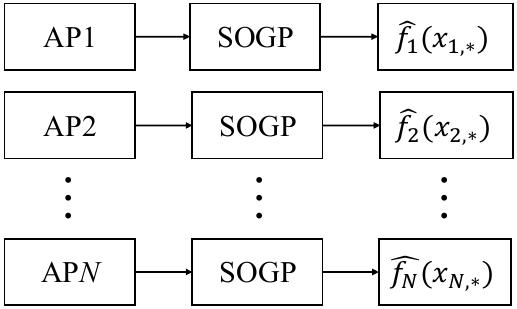}\\
    \vspace{0.1cm}
    {\scriptsize (a)}\\
    \includegraphics[angle=-90,width=.6\linewidth]{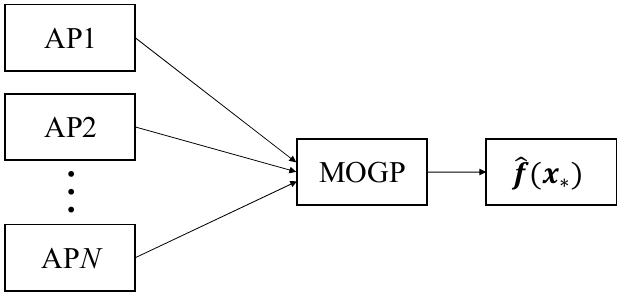}\\
    {\scriptsize (b)}
  \end{center}
  \caption{RSSI data augmentation based on (a) SOGP and (b) MOGP.}
  \label{fig:gp-data-augmentation}
\end{figure}
During the construction of RSSI dataset, RSSIs from multiple APs (i.e., multiple
outputs) are measured simultaneously at a given reference point (i.e., an
input). As shown in Fig.~\ref{fig:gp-data-augmentation}~(a), we can fit multiple
SOGP models, i.e., one for each AP by separately and independently processing
the RSSI data from that AP. In this case, we can exploit only the spatial
correlation among them. To exploit the correlation among RSSIs from multiple
APs, especially the APs deployed closely to each other, we process them in an
integrated way using a single MOGP as shown in
Fig.~\ref{fig:gp-data-augmentation}~(b).


\section{RSSI Data Augmentation Based on MOGP}
\label{sec:rssi-da-mogp}
\begin{figure}[!tbh]
  \centering%
  \includegraphics[width=\linewidth]{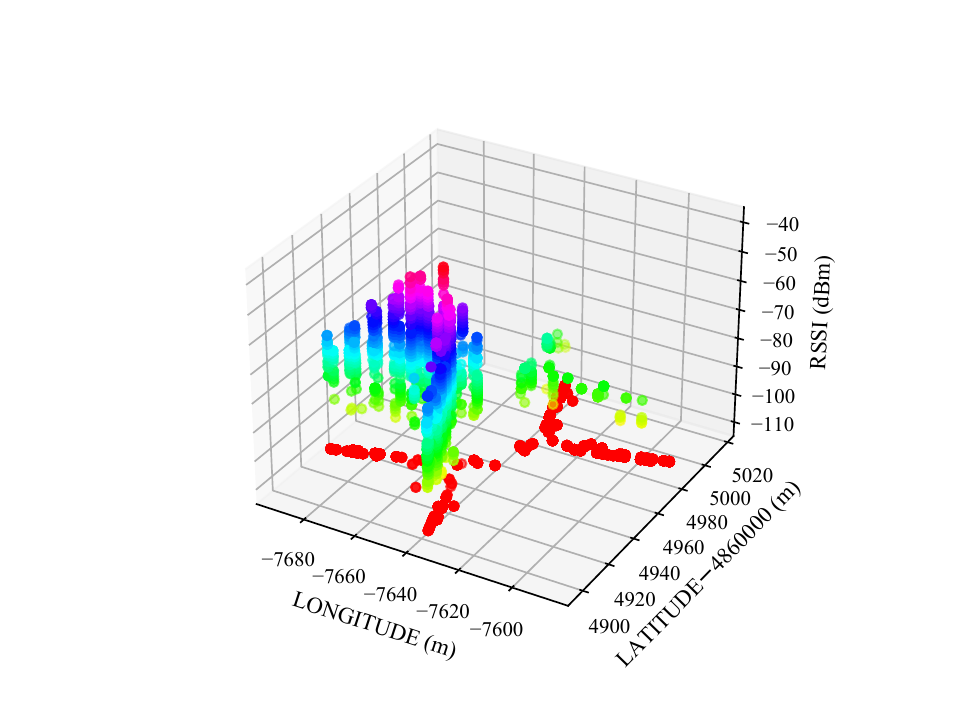}
  \caption{RSSI measurements for WAP13 at building~0 and floor~0 of the
    UJIIndoorLoc database~\cite{UJI}.}
  \label{fig:data}
\end{figure}

Figure~\ref{fig:data} shows the RSSI samples for WAP13 at building 0 and floor 0
of the UJIIndoorLoc database.
The red points on the bottom of Fig.~\ref{fig:data} are the reference points
where RSSIs were measured, whose cross shape is consistent with the actual shape
of the building (i.e., Fig.~4 of~\cite{UJI}). The colors of the points off the
bottom represent their signal strengths. When the details of the floor plan are
not available, we have to rely on the extreme coordinates of the reference
points (i.e., upper left and lower right points based on the assumption of a
rectangular floor) during the sampling, which, however, could affect the
localization accuracy as discussed in
Section~\ref{sec:experiment-results}. Based on the investigation of the RSSI
statistics of the UJIIndoorLoc database, we consider three different methods of
data augmentation by MOGP, which are discussed in
Sections~\ref{sec:da-single-floor} to \ref{sec:da-single-building}.

\subsection{By A Single Floor}
\label{sec:da-single-floor}
\begin{figure}[!tbh]
  \centering%
  \includegraphics[angle=-90,width=.6\linewidth]{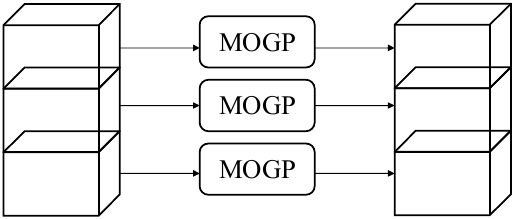}
  \caption{Data augmentation by a single floor.}
  \label{fig:da-single-floor}
\end{figure}

This augmentation method is the simplest of all because it fits an MOGP based
only on the RSSIs from APs on a single floor; sampling for the fake RSSI
generation is also limited to the same floor as shown in
Fig.~\ref{fig:da-single-floor}.
This method is suitable for a floor with lower signal attenuation in the
horizontal direction but higher signal attenuation in the vertical direction,
the latter of which reduces the effect of APs located on different floors on the
RSSIs on the floor under consideration. Unlike SOGP limited to one AP in
fitting, this method still can take into account the effect of all the APs on
the same floor.

\subsection{By Neighboring Floors}
\label{sec:da-neighboring-floors}
\begin{figure}[!tbh]
  \centering%
  \includegraphics[angle=-90,width=.6\linewidth]{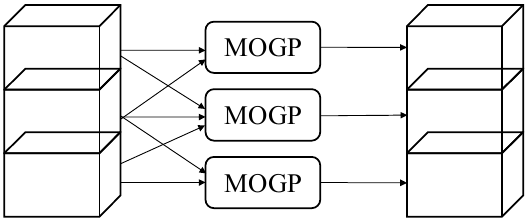}
  \caption{Data augmentation by neighboring floors.}
  \label{fig:da-neighboring-floors}
\end{figure}

When there exist impacts on RSSIs from APs on different floors, which is the
case for the UJIIndoorLoc database with the floor height of \SI{4}{\m}, we can
extend the two-dimensional correlation of the augmentation method \textit{by a
  single floor} to the three-dimensional one as shown
in~Fig.~\ref{fig:da-neighboring-floors} by including RSSIs of the neighboring
floors when aggregating the data for the current floor.
For example, when a target is the second floor, we fit an MOGP model to the data
from the first and the third floors as well as those of the second floor.
Compared with the data augmentation method by a single floor, one significant
difference is that there is one additional dimension in the kernel to contain
height or floor information.

\subsection{By A Single Building}
\label{sec:da-single-building}
\begin{figure}[!tbh]
  \centering%
  \includegraphics[angle=-90,width=.6\linewidth]{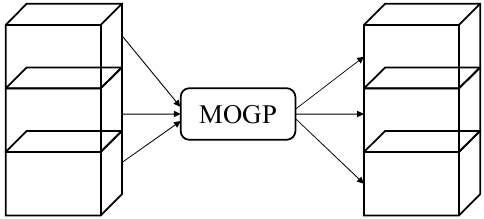}
  \caption{Data augmentation by a Single building.}
  \label{fig:da-single-building}
\end{figure}

In this method, the data of a building is taken as a whole in fitting one common
MOGP model as shown in Fig.~\ref{fig:da-single-building}.
When a building contains staircases, patios, lofts, and so on, this method can
consider the correlation of signals in those special structures depending on the
coverage of the original data.

\section{Experiment Results}
\label{sec:experiment-results}
We select the state-of-the-art RNN model with Long Short-Term Memory (LSTM)
cells~\cite{2021hierarchical} as our reference localization scheme and carry out
experiments with the original and the augmented UJIIndoorLoc datasets in order
to investigate the effects of the MOGP-based augmentation on the localization
performance. The MOGP regression is implemented based on GPy---i.e., a GP
framework in Python developed at the Sheffield machine learning
group~\cite{GPy}---following the steps outlined in Section \ref{sec:math}.
The Stacked AutoEncoder (SAE) of the RNN model consists of three hidden layers
of 256, 128, and 64 nodes, which is followed by two common hidden layers with
128 and 128 nodes. For building and floor classifiers, we have two stacked LSTM
cells followed by two Fully-Connected (FC) layers of 32 nodes and 1 output
node. Position estimator consists of three FC layers of 512 and 512 nodes and 2
output nodes for two-dimensional location coordinates. We apply \textit{early
  stopping} with patience of 10 for coordinate estimation model and 30 with
\textit{save best only} functions activated for building/floor classification
model. Table~\ref{tbl:hyperSetting} summarizes the key RNN parameter values for
the experiments.
\begin{table}[!tbh]
  \caption{RNN parameter values.}
  \label{tbl:hyperSetting}
  \centering%
  \begin{tabular}{ll}
    \hline
    Parameter                   & Value      \\
    \hline
    SAE Hidden Layers           & 256-128-64 \\
    SAE Activation              & ReLu       \\
    SAE Optimizer               & Adam       \\
    SAE Loss                    & MSE        \\
    Common Hidden Layers        & 128-128    \\
    Common Activation           & ReLu       \\
    Common Dropout              & 0.2        \\
    Common Loss                 & MSE        \\
    LSTM Cells                  & 256-256    \\
    LSTM Activation             & ReLu       \\
    LSTM Optimizer              & Adam       \\
    LSTM Loss                   & MSE        \\
    BF Classifier Hidden Layers & 32-1       \\
    BF Classifier Activation    & MSE        \\
    BF Classifier Optimizer     & Adam       \\
    BF Classifier Dropout       & 0.2        \\
    BF Classifier Loss          & ReLu       \\
    Position Hidden Layers      & 512-512-2  \\
    Position Activation         & MSE        \\
    Position Optimizer          & Adam       \\
    Position Dropout            & 0.1        \\
    Position Loss               & tanh       \\
    \hline
  \end{tabular}
\end{table}

The localization performance of the RNN model with the original and augmented
UJIIndoorLoc datasets is summarized in Table~\ref{tbl:result_comp}, where the
floor hit rate is for the correct identification of both building and floor
Identifiers (IDs) and the 3D error is the mean of three-dimensional Euclidean
distances~\cite{EvAAL}.
\begin{table*}[!tbh]
  \caption{Comparison of the localization performance with the original and the
    augmented UJIIndoorLoc~\cite{UJI} datasets.}
  \label{tbl:result_comp}
  \centering%
  \begin{tabular}{|l|r|r|r|r|}
    \hline
    \multicolumn{1}{|c|}{\multirow{2}{*}{Performance~metric}}    & \multicolumn{1}{c|}{\multirow{2}{*}{Original~dataset}} &
                                                                                                                       \multicolumn{3}{c|}{Augmented~dataset} \\ \cline{3-5}
                                                                 & &
                                                                     By~a~single~floor & By~neighboring~floors & By~a~single~building \\ \hline
    Building hit rate [\%] & \textbf{99.99}       & \textbf{99.99} & 99.80    & \textbf{99.99}    \\ \hline
    Floor hit rate [\%]    & 92.92       & 93.12 & 92.70    & \textbf{93.84}    \\ \hline
    3D error [\si{m}]   & 8.61        & 8.87  & 8.85     & \textbf{8.59}     \\ \hline
  \end{tabular}
\end{table*}
It is clear from the results that the data augmentation by a single building
outperforms the other two augmentation methods as well as the original data; for
all three performance metrics, the data augmentation by a single building
provides the best results, though the differences are minor. As discussed in
Section~\ref{sec:rssi-da-mogp}, the data augmentation by a single building can
take into account the correlation among all RSSI data of a building and thereby
provide better predictions at sampling points. These results show that the
effects of APs located on different floors---including those on non-neighboring
floors---on RSSIs cannot be ignored for the UJIIndoorLoc dataset.

We also compare the localization performance of the RNN with the UJIIndoorLoc
dataset augmented by a single building with the three best results from the
EvAAL competition~\cite{EvAAL} as well as the results of~\cite{2021hierarchical}
in Table~\ref{tbl:result_comp_o}.
\begin{table*}[!tbh]
  \caption{Result comparison}
  \label{tbl:result_comp_o}
  \centering%
  \begin{tabular}{|c|c|c|c|c|c|c|}
    \hline
    Performance metric & \textbf{RNN with augmented dataset} & RNN~\cite{2021hierarchical} & MOSAIC & HFTS  & ICSL  \\ \hline
    Building hit rate [\%] & \textbf{100}   & \textbf{100}      & 98.65  & \textbf{100}   & \textbf{100}   \\ \hline
    Floor hit rate [\%]   & 94.20 & 95.23    & 93.86  & \textbf{96.25} & 86.93 \\ \hline
    3D error [\si{\m}]        & \textbf{8.42}  & 8.62     & 11.64  & 8.49  & 7.67  \\ \hline
  \end{tabular}
\end{table*}
Note that the comparison in Table~\ref{tbl:result_comp_o} is not done on a fair
and strict basis: The original UJIIndoorLoc testing set is not publicly
available but was given to those participating in the EvAAL competition; the
results from the experiments in this paper and those in~\cite{2021hierarchical}
are based only on the training and the validation sets of the UJIIndoorLoc
database, where the validation set was splitted into new validation and testing
sets. Also, note that the results in Table~\ref{tbl:result_comp_o} is the best
ones unlike the average ones in~\ref{tbl:result_comp}. Still, the results in
Table~\ref{tbl:result_comp_o} are enough to demonstrate the feasibility of the
proposed RSSI augmentation by a single building, where the 3D positioning error
is reduced by \SI{0.2}{\m} compared to that of~\cite{2021hierarchical}.

\section{Concluding Remarks}
\label{sec:concluding-remarks}
In this paper, we have proposed MOGP-based RSSI data augmentation for
multi-building and multi-floor indoor localization and evaluated its
localization performance using the RNN model~\cite{2021hierarchical} and the
UJIIndoorLoc dataset~\cite{UJI}. We first compared three different methods of
augmentation---i.e., \textit{by a single floor}, \textit{by neighboring floors},
and \textit{by a single building}---and found that the augmentation by a single
building outperforms the other two. Then, we also compared the localization
performance of the RNN model with the UJIIndoorLoc data augmented by a single
floor with the three best results from the EvAAL competition~\cite{EvAAL} as
well as the results of~\cite{2021hierarchical}.
These results demonstrate the feasibility of using MOGP to augment the existing
localization databases, which is important given the difficulty of data
collection in an indoor environment, especially under the current epidemic
situation of COVID-19.

Note that the localization error can be further reduced if the database can
provide more detailed information on floor plans:
When generating fake RSSI data with MOGP, sampling locations based on the simple
assumption of rectangular floors could be located outside actual buildings like
the cross-shaped building in the UJIIndoorLoc database discussed in
Section~\ref{sec:rssi-da-mogp}. In such a case, we can adjust sampling points
based on the detailed floor plans from the database, e.g., imposing more
limiting parameters on MOGP or filtering sampling points based on the floor
plans, which is to be investigated in future work.

\section*{Acknowledgment}
This work was supported in part by Postgraduate Research Scholarship (under
Grant PGRS1912001) and Key Program Special Fund (under Grant KSF-E-25) of Xi'an
Jiaotong-Liverpool University.%


\begin{thebibliography}{10}
\providecommand{\url}[1]{#1}
\csname url@samestyle\endcsname
\providecommand{\newblock}{\relax}
\providecommand{\bibinfo}[2]{#2}
\providecommand{\BIBentrySTDinterwordspacing}{\spaceskip=0pt\relax}
\providecommand{\BIBentryALTinterwordstretchfactor}{4}
\providecommand{\BIBentryALTinterwordspacing}{\spaceskip=\fontdimen2\font plus
\BIBentryALTinterwordstretchfactor\fontdimen3\font minus
  \fontdimen4\font\relax}
\providecommand{\BIBforeignlanguage}[2]{{%
\expandafter\ifx\csname l@#1\endcsname\relax
\typeout{** WARNING: IEEEtran.bst: No hyphenation pattern has been}%
\typeout{** loaded for the language `#1'. Using the pattern for}%
\typeout{** the default language instead.}%
\else
\language=\csname l@#1\endcsname
\fi
#2}}
\providecommand{\BIBdecl}{\relax}
\BIBdecl

\bibitem{survey}
F.~Zafari, A.~Gkelias, and K.~K. Leung, ``A survey of indoor localization
  systems and technologies,'' \emph{{IEEE} Commun. Surveys Tuts.}, vol.~21,
  no.~3, pp. 2568--2599, 2019.

\bibitem{zhenghang18:_xjtluin}
Z.~Zhong, Z.~Tang, X.~Li, T.~Yuan, Y.~Yang, W.~Meng, Y.~Zhang, R.~Sheng,
  N.~Grant, C.~Ling, X.~Huan, K.~S. Kim, and S.~Lee, ``{XJTLUIndoorLoc}: A new
  fingerprinting database for indoor localization and trajectory estimation
  based on {Wi-Fi} {RSS} and geomagnetic field,'' in \emph{Proc. {CANDAR}'18},
  Hida Takayama, Japan, Nov. 2018.

\bibitem{DNN}
\BIBentryALTinterwordspacing
K.~S. Kim, S.~Lee, and K.~Huang, ``A scalable deep neural network architecture
  for multi-building and multi-floor indoor localization based on {Wi-Fi}
  fingerprinting,'' \emph{Big Data Analytics}, vol.~3, no.~1, Apr 2018.
  [Online]. Available: \url{http://dx.doi.org/10.1186/s41044-018-0031-2}
\BIBentrySTDinterwordspacing

\bibitem{CNN1}
X.~Song, X.~Fan, X.~He, C.~Xiang, Q.~Ye, X.~Huang, G.~Fang, L.~L. Chen, J.~Qin,
  and Z.~Wang, ``{CNNLoc}: Deep-learning based indoor localization with {WiFi}
  fingerprinting,'' in \emph{Proc. 2019 {IEEE} SmartWorld, Ubiquitous
  Intelligence Computing, Advanced Trusted Computing, Scalable Computing
  Communications, Cloud Big Data Computing, Internet of People and Smart City
  Innovation ({SmartWorld/SCALCOM/UIC/ATC/CBDCom/IOP/SCI})}, 2019, pp.
  589--595.

\bibitem{CNN2}
M.~Ibrahim, M.~Torki, and M.~ElNainay, ``{CNN} based indoor localization using
  {RSS} time-series,'' in \emph{Proc. 2018 {IEEE} Symposium on Computers and
  Communications ({ISCC})}, 2018, pp. 01\,044--01\,049.

\bibitem{2021hierarchical}
A.~E.~A. Elesawi and K.~S. Kim, ``Hierarchical multi-building and multi-floor
  indoor localization based on recurrent neural networks,'' 2021.

\bibitem{UJI}
J.~Torres-Sospedra, R.~Montoliu, A.~Martínez-Usó, J.~P. Avariento, T.~J.
  Arnau, M.~Benedito-Bordonau, and J.~Huerta, ``Ujiindoorloc: A new
  multi-building and multi-floor database for {WLAN} fingerprint-based indoor
  localization problems,'' in \emph{Proc. 2014 International Conference on
  Indoor Positioning and Indoor Navigation ({IPIN})}, 2014, pp. 261--270.

\bibitem{rasmussenGaussianProcessesMachine2006}
C.~E. Rasmussen and C.~K.~I. Williams, \emph{Gaussian processes for machine
  learning}, ser. Adaptive Computation and Machine Learning.\hskip 1em plus
  0.5em minus 0.4em\relax {Cambridge, MA}: {MIT Press}, 2006.

\bibitem{Kriging}
\BIBentryALTinterwordspacing
S.-S. Jan, S.-J. Yeh, and Y.-W. Liu, ``Received signal strength database
  interpolation by {Kriging} for a {Wi-Fi} indoor positioning system,''
  \emph{Sensors}, vol.~15, no.~9, pp. 21\,377--21\,393, 2015. [Online].
  Available: \url{https://www.mdpi.com/1424-8220/15/9/21377}
\BIBentrySTDinterwordspacing

\bibitem{njimaIndoorLocalizationUsing2021}
W.~Njima, M.~Chafii, A.~Chorti, R.~M. Shubair, and H.~V. Poor, ``Indoor
  localization using data augmentation via selective generative adversarial
  networks,'' \emph{{IEEE} Access}, vol.~9, pp. 98\,337--98\,347, 2021.

\bibitem{intro_kernel}
\BIBentryALTinterwordspacing
F.~Jäkel, B.~Schölkopf, and F.~A. Wichmann, ``A tutorial on kernel methods
  for categorization,'' \emph{Journal of Mathematical Psychology}, vol.~51,
  no.~6, pp. 343--358, 2007. [Online]. Available:
  \url{https://www.sciencedirect.com/science/article/pii/S0022249607000375}
\BIBentrySTDinterwordspacing

\bibitem{mean}
\BIBentryALTinterwordspacing
E.~Schulz, M.~Speekenbrink, and A.~Krause, ``A tutorial on gaussian process
  regression: Modelling, exploring, and exploiting functions,'' \emph{Journal
  of Mathematical Psychology}, vol.~85, pp. 1--16, 2018. [Online]. Available:
  \url{https://www.sciencedirect.com/science/article/pii/S0022249617302158}
\BIBentrySTDinterwordspacing

\bibitem{MOGPremarks}
\BIBentryALTinterwordspacing
H.~Liu, J.~Cai, and Y.-S. Ong, ``Remarks on multi-output gaussian process
  regression,'' \emph{Knowledge-Based Systems}, vol. 144, pp. 102--121, 2018.
  [Online]. Available:
  \url{https://www.sciencedirect.com/science/article/pii/S0950705117306123}
\BIBentrySTDinterwordspacing

\bibitem{bruinsmaScalableExactInference}
W.~P. Bruinsma, E.~Perim, W.~Tebbutt, J.~S. Hosking, A.~Solin, and R.~E.
  Turner, ``Scalable exact inference in multi-output {Gaussian} processes,'' in
  \emph{Proc. the 37th International Conference on Machine Learning}, ser.
  Proceedings of Machine Learning Research, vol. 119.\hskip 1em plus 0.5em
  minus 0.4em\relax {PMLR}, July 2020, pp. 1190--1201.

\bibitem{kernel_calculation}
\BIBentryALTinterwordspacing
D.~Duvenaud, J.~Lloyd, R.~Grosse, J.~Tenenbaum, and G.~Zoubin, ``Structure
  discovery in nonparametric regression through compositional kernel search,''
  vol.~28, no.~3, pp. 1166--1174, 17--19 Jun 2013. [Online]. Available:
  \url{https://proceedings.mlr.press/v28/duvenaud13.html}
\BIBentrySTDinterwordspacing

\bibitem{RBF}
M.~Kanagawa, P.~Hennig, D.~Sejdinovic, and B.~K. Sriperumbudur, ``Gaussian
  processes and kernel methods: A review on connections and equivalences,''
  2018.

\bibitem{Matern}
M.~L. Stein, \emph{Interpolation of spatial data: some theory for
  kriging}.\hskip 1em plus 0.5em minus 0.4em\relax Springer Science \& Business
  Media, 1999.

\bibitem{GPy}
\BIBentryALTinterwordspacing
{GPy} - a {Gaussian} process (gp) framework in {Python}. [Online]. Available:
  \url{https://gpy.readthedocs.io/en/deploy/}
\BIBentrySTDinterwordspacing

\bibitem{EvAAL}
A.~Moreira, M.~J. Nicolau, F.~Meneses, and A.~Costa, ``{Wi-Fi} fingerprinting
  in the real world - {RTLS@UM} at the {EvAAL} competition,'' in \emph{Proc.
  2015 International Conference on Indoor Positioning and Indoor Navigation
  ({IPIN})}, 2015, pp. 1--10.

\end{thebibliography}

\end{document}